\documentclass[12pt]{article}
\usepackage{url}
\usepackage{vmargin,bm,dsfont,cleveref,amsmath,graphics,graphicx,algorithm,algpseudocode,subfloat,subfig,authblk}
\setpapersize{USletter}
\setmarginsrb{1in}{1in}{1in}{0.5in}{0pt}{0mm}{10pt}{0.5in}
\usepackage{color,xcolor}
\def\argmin{\mathop{\rm argmin}}

\begin{document}

\title{Point Spread Function Engineering for 3D Imaging of Space Debris using a Continuous Exact $\ell_0$ Penalty (CEL0) Based Algorithm\thanks{The research of the first, third and fourth authors was supported by the US Air Force Office of Scientific Research under grant FA9550-15-1-0286. The work of the second author was supported by HKRGC Grant CUHK14306316, HKRGC Grant CUHK14301718, HKRGC CRF Grant C1007-15G, HKRGC AoE Grant AoE/M-05/12, CUHK DAG No. 4053211, and CUHK FIS Grant No. 1907303. 
The work of the third author was also supported by HKRGC Grant CUHK14306316.  }}

\author[1]{Chao Wang}
\author[1]{Raymond H. Chan}
\author[2]{Robert J. Plemmons}
\author[3,4]{Sudhakar Prasad}
\affil[1]{Department of Mathematics, The Chinese University of Hong Kong, Shatin, Hong Kong}
\affil[2]{Departments of Mathematics and Computer Science, Wake Forest University, NC 27109, USA}
\affil[3]{Department of Physics and Astronomy, The University of New Mexico, NM 87131, USA}
\affil[4]{School of Physics and Astronomy, University of Minnesota, Minneapolis, MN 55455, USA}
\date{ }

\maketitle

\begin{abstract}
 We consider three-dimensional (3D) localization and imaging  of space debris from only one two-dimensional (2D) snapshot image. The technique involves an optical imager that exploits off-center image rotation to encode both the lateral and depth coordinates of point sources, with the latter being encoded in the angle of rotation of the PSF. We formulate 3D localization into a large-scale sparse 3D inverse problem in discretized form.  A recently developed penalty called  continuous exact $\ell_0$ (CEL0)  is applied in this problem for the Gaussian noise model. Numerical experiments and comparisons illustrate the efficiency of the algorithm.

  \end{abstract}

\section{Introduction}
\label{sec:intro}  

The area of 3D imaging and localization  
has been getting increasing attention in recent years. The use of 3D localization in single-molecule super-resolution microscopy can obtain a complete picture of subcellubar structures   \cite{Book_SR_micro2017,tilted20183d, 3dsml2017review}. The molecules are labeled by some specific fluorescent proteins or oligonucleotides, which can be regarded as a collection of point sources. Another application of 3D imaging is for space situational awareness (SSA). Currently, there are more than 20,000 objects in orbit around earth \cite{nature_news2018quest}, including operational satellites, dead ones and other human-made debris.  3D localization of micro-scale space debris that become increasingly abundant with decreasing size can be vital for SSA systems responsible for the overall protection of space assets.  Radar systems can sometimes detect such space debris objects, but can at best localize them with lower precision than short-wavelength optical systems. 
A stand-alone optical system based on the use of a light-sheet illumination and scattering concept \cite{englert2014optical}
for spotting debris within meters of a spacecraft has also been proposed.  
A second system can localize all three coordinates of an unresolved, scattering debris \cite{hampf2015optical,wagner2016detection} by
utilizing either the parallex between two observations, or a pulsed laser ranging system, or a hybrid system.
 However, to the best of our knowledge there is no other proposal of either an optical or an integrated optical-radar system to perform full 3D debris localization and tracking in the range of tens to hundreds of meters. 
Prasad \cite{Internal_report2016prasad} has proposed engineering point spread functions for 3D localization by
the use of an optical imager that exploits off-center image rotation.  This system encodes in a single image snapshot both the range $z$ and transverse ($x,y$) coordinates of a swarm of unresolved sources such as 
small, sub-centimeter class space debris, which when actively illuminated can scatter a fraction of laser irradiance back into the imaging sensor.  2D image data taken with a specially designed point spread function (PSF) that encodes, via a simple rotation, changing source distance can be employed to acquire a three dimensional (3D) field of unresolved sources like space debris. Here, we propose the 3D localization and tracking of space debris at optical wavelengths by PSF engineering and employing a space-based telescope.

 PSF engineering is widely used in single-molecule super-resolution \cite{lew2011corkscrew,DH2008pavani,DH2009pavani,tetrapods2014shechtman,tetrapods2015shechtman,Rice2016generalized}. It is based on choosing a phase pattern that makes the defocused image of a point source depth-dependent without blurring it excessively. For space-surveillance and SSA, PSF engineering is just beginning to be considered, with \cite{kumar2013psf,prasad2013rotating} proposing the optical theory and simulation of a single lobe rotating PSF. Such PSFs have obvious advantages over multi-lobe PSFs when dealing with high source densities at low light levels. In \cite{kl_nc}, we proposed a mathematical formulation of the 3D localization problem employing such a PSF in the Poisson-noise case. This noise model characterizes an EMCCD sensor operated in the photon-counting (PC) regime. However, when conventional CCD sensors operate at low per-pixel photon fluxes and large read-out noise, a Gaussian noise model describes more accurately this kind of data noise.  In this paper, we consider the latter case. 

The following forward model based on the rotating PSF image describes the spatial distribution of image brightness for $M$ point sources in the observed 2D image: 
\begin{equation}
	G (x,y) = \sum_{i=1}^M \mathcal{H}_{{z_i}}(x-x_i,y-y_i)f_i + b + \mathcal{N}(x,y), 
	\label{equ:forward_model}
\end{equation}
where  $\mathcal{N}$ is the Gaussian noise operator which is data-independent and $b$ is the uniform background value. Here  
$\mathcal{H}_{z_i}(x-x_i,y-y_i)$ is the rotating PSF for the $i$-th point source of flux $f_i$ and  3D position coordinates $(x_i, y_i, z_i)$ with the depth information $z_i$ encoded in  $\mathcal{H}_{z_i}$, and $(x, y)$ is  the position in the image plane.  
In the Fourier optics model \cite{Goodman17} of image formation, the  incoherent PSF for a clear aperture containing a phase mask that imposes an optical phase retardation, $\psi(\mathbf{s})$, on the imaging wavefront is given by 
\begin{equation}
	\label{equ:A}
{\mathcal{H}_{z}}(\mathbf{s} ) = 
{1\over \pi}\left|\int P(\mathbf{u} )\mathrm{exp} \left[ \iota( 2\pi \mathbf{u}\cdot\mathbf{s} +  \zeta u^2 - \psi(\mathbf{u}))  \right] d \mathbf{u} \right|^2,
\end{equation}
where 
$\zeta = \frac{\pi  (l_0-z) R^2}{\lambda l_0 z}$  is defocus parameter and the imaging wavelength is denoted by $\lambda$.  Here $\iota = \sqrt{-1} $, $\l_0$ is  the distance between the lens and the best focus point, and $P(\mathbf{u} )$ is  the indicator function for the pupil of radius $R$. We use $\mathbf{s}$ with polar coordinates $(s, \phi_{\mathbf{s}}) $ to denote a scaled version of the image-plane position vector, $\mathbf{r}$, namely $\mathbf{s} =  \frac{\mathbf{r}}{\lambda z_I/R} $.  
Here $\mathbf{r}$ is measured from the center of the geometric {(Gaussian)} image point located at $\mathbf{r}_I=(x,y)$, and $z_I$ is the distance between the image plane and the lens. The pupil-plane position vector $\boldsymbol{\rho}$ is normalized by the pupil radius, $\mathbf{u} = \frac{\boldsymbol{\rho}}{R}$. 
For the single-lobe rotating PSF, $\psi(\mathbf{u}) $ is chosen to be the spiral phase
 profile defined as $$\psi(\mathbf{u}) = l\phi_{\mathbf{u}}, \ \ \text{for } \sqrt{\frac{l-1}{L}}\leq u\leq \sqrt{\frac{l}{L}}, \ l = 1,\cdot\cdot\cdot, L, $$  in which $L$ is the number of concentric annular zones in the phase mask. We evaluate \eqref{equ:A} by using the fast Fourier transform. With such spiral phase retardation, PSF (\ref{equ:A}) performs a complete rotation about the geometrical image center, as $\zeta$ changes between $-L\pi$ and $L\pi$, before it begins to degrade significantly for values of $\zeta$ outside this range.

		Next, we discuss  the problem of 3D localization of closely spaced point sources from simulated noisy image data obtained by using such a rotating-PSF imager. The localization problem is discretized on a cubical lattice where the coordinates and values of its nonzero entries represent the 3D locations and fluxes of the sources, respectively. Finding the locations and fluxes of a few point sources on a large lattice is evidently a large-scale sparse 3D inverse problem. Based on the Gaussian statistical noise model, we describe the results of simulation using a recently developed regularization tool called the continuous exact $\ell_0$ (CEL0) penalty term \cite{CEL02015soubies}, which when added to a least-squares data fitting term constitutes an $\ell_0$-sparsity non-convex optimization protocol with promising results. We use an iteratively reweighted $\ell_1$ (IRL1) algorithm to solve this optimization problem.

The rest of the paper is organized as follows. In Section \ref{sec:non-convex}, we  describe the CEL0-based non-convex optimization model for solving the point source localization problem. In Section \ref{sec:alg}, our non-convex optimization algorithm is developed. Numerical experiments, including comparisons with other optimization methods, are discussed in Section \ref{sec:numerical}.  Some concluding remarks are made in Section \ref{sec:Conclusions}.

\section{CEL0-based Optimization Model}\label{sec:non-convex}

Here, we  build  a forward model for the problem based  on the approach developed in  \cite{Rice2016generalized}. 
In order to estimate the 3D locations of the point sources, we assume {that they are distributed on } 
a discrete lattice  $\mathcal{X}\in \mathds{R}^{m \times n \times d}$. 
The indices of the nonzero entries of $\mathcal{X}$ are the 3-dimensional locations of the point sources and the values at these entries   correspond to the fluxes, {\it i.e.}, the energy emitted by the illuminated point source. The observed 2D image $G \in \mathds{R}^{m \times n}$ can be approximated  as 
\begin{equation*}
	G \approx \mathcal{T}(\mathcal{A} \ast \mathcal{X} )+b 1 + \mathcal{N},
\end{equation*}
where
 $b$ is background signal, $1$ is a matrix of $1$s of size the same as the size of $G$ and 
$\mathcal{N}$ is the Gaussian noise. Here,   $\mathcal{A} \ast \mathcal{X}$ is the convolution of $\mathcal{X}$ with the 3D PSF $\mathcal{A}$. This 3D PSF ${\mathcal{A}}$ is a cube which is constructed by a sequence of images with respect to different depths of the points. Each horizontal slice is the image corresponding to a point source at the origin in the $(x,y)$ plane and at depth $z$.  This tensor $\mathcal{A}$ {is constructed} by sampling depths at regular intervals in the range, $\zeta_i\in [-\pi L, \ \pi L]$, over which the PSF performs one complete rotation about the geometric image center before it begins to break apart. The $i$-th slice of the dictionary is {$\mathcal{H}_{z_i}$ with certain depth $z_i$}.  Here $\mathcal{T}$ is an operator for extracting the last slice of the cube $\mathcal{A} \ast \mathcal{X}$ since the observed information is a snapshot.

\begin{figure}[htbp]
\centering
\includegraphics[width=0.46\textwidth, height=0.27\textheight]{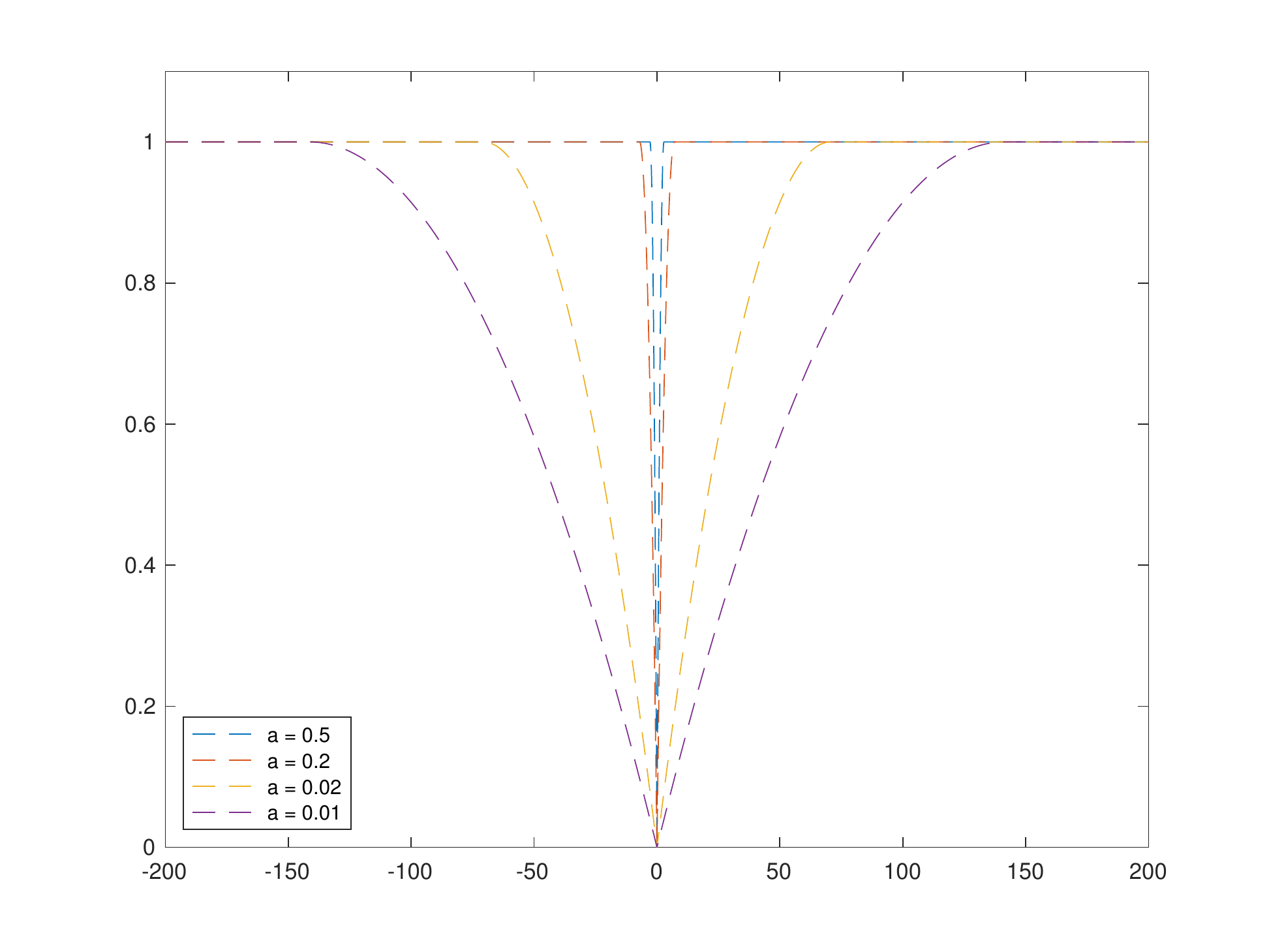}
\caption{The function $\phi(a, \mu; u)$ for $\ell_2$-CEL0  with $\mu=1$.  } 
\label{fig:curve_regul}
\end{figure}

 In order to recover $\mathcal{X}$,  we need to solve a large-scale sparse 3D inverse problem with data-fitting term and regularization term.  
Since the Gaussion noise is data-independent, it leads to the use of  least squares for the data-fitting term, {\it i.e.,} $ \frac{1}{2} \left\|\mathcal{T}(\mathcal{A}\ast \mathcal{X})+b 1 -G\right\|_F^2, $
where  $\|Y\|_F$ is the Frobenius norm of $Y$, which is equal to the $\ell_2$ norm of the vectorized $Y$. 
For the regularization term, we choose the continuous exact $\ell_0$ (CEL0) penalty, as described in  \cite{cel0_cell2017high,CEL02015soubies,cel0_theory2017unified}.  It is a non-convex term approaching the $\ell_0$ 
 norm for linear least squares data fitting problems and is constructed as
 $$\mathcal{R}(\mathcal{X}):=\Phi_{\mathrm{CEL0}}(\mathcal{X})=\sum_{u,v,w = 1}^{m,n,d}\phi(\|\mathcal{T}(\mathcal{A} \ast \delta_{uvw} )\|_F, \mu; \mathcal{X}_{uvw}), $$
  where $\phi(a, \mu; u) = \mu  - \frac{a^2}{2}\left(|u|-\frac{\sqrt{2\mu}}{a}\right)^2 \mathds{1}_{\left\{|u| \leq \frac{\sqrt{2 \mu  }}{a} \right\}}$; see Figure~\ref{fig:curve_regul}, and
$\mathds{1}_{\{ u \in E\}} := \begin{cases}
	1 & \text{if} \ u\in E;\\
	0 & \text{others}.
\end{cases}   $ 
Here $\delta_{uvw}$ is a 3D tensor whose only nonzero entry is at $(u, v, w)$ with value 1 and $\mu $ is the regularization parameter.

The minimization problem may be stated as
	 \begin{equation}
	 	\min\limits_{\mathcal{X}\geq 0 }\left\{  \frac{1}{2} \left\|\mathcal{T}(\mathcal{A}\ast \mathcal{X})+b-G\right\|_F^2 +\sum_{u,v,w = 1}^{m,n,d}\phi(\|\mathcal{T}(\mathcal{A} \ast \delta_{uvw} )\|_F,\mu; \mathcal{X}_{uvw}) \right\}. \label{equ:min_fun_gaussian}
	 \end{equation}
	 To emphasize that our non-convex optimization model is based on the CEL0 regularization term, we simply designate our optimization model \eqref{equ:min_fun_gaussian} as CEL0.

When combined with a least-squares data fitting term, CEL0 has many good properties and it does not place any strict requirements on the former.
{The global minimizers of the $\ell_0$ penalty model with a least squares data-fitting term ($\ell_2$-$\ell_0$)  are, in fact, contained in the set of global minimizers of CEL0 \eqref{equ:min_fun_gaussian}. A minimizer of \eqref{equ:min_fun_gaussian} can be transformed into a minimizer of  $\ell_2$-$\ell_0$. Moreover, some local minimizers of $\ell_2$-$\ell_0$ are not critical points of CEL0, which means CEL0 can avoid some local minimizers of $\ell_2$-$\ell_0$.
}

\section{ Development of the Algorithm  }\label{sec:alg}

 Note that our optimization model for the Gaussian noise case is  non-convex, due to the regularization term. 
We first consider an iterative reweighted $\ell_1$ algorithm (IRL1) \cite{IRL1_2015} to solve the optimization problem. This is a majorization-minimization method which solves a series of convex optimization problems with a weighted-$\ell_1$ regularization term. 
 It considers the problem (see Algorithm 3, in \cite{IRL1_2015})
\begin{equation*}
	\min_{x\in X} F(x):=F_1(x)+ F_2(G(x)),
\end{equation*}
 where $X$ is the constraint set, $F$ is a lower semicontinuous (lsc) function, extended, real-valued, proper, while $F_1$ is proper, lower-semicontinous, and convex and $F_2$ is coordinatewise nondecreasing, {\it i.e.} $F_2(x)\leq F_2(x+t e_i)$ with $x, x+t e_i\in G(X)$ and $t >0,$ where  $e_i$ is the $i$-th canonical basis unit vector. The function $F_2$ is concave on $G(X)$.   The IRL1 iterative scheme \cite[Algorithm 3]{IRL1_2015} is 
 \begin{equation*}
 	\begin{cases}
 		W^{(k)} = \partial F_2(y), \ y = G(x^{(k)}), \\
 		x^{(k+1)} = \argmin\limits_{x\in X} \left\{F_1(x) + \langle W^{(k)}, G(x)\rangle \right\},
 	\end{cases}
 \end{equation*}
where $\partial$ stands for subdifferential.

For the Gaussian noise problem \eqref{equ:min_fun_gaussian}, we choose
\begin{equation*}
	\begin{split}
		F_1(\mathcal{X}) = & \frac{1}{2} \left\|\mathcal{T}(\mathcal{A} \ast \mathcal{X} )+b 1 -G\right\|_F^2;   \\
		F_2(\mathcal{X}) = & \  \mu  - \frac{a_i^2}{2}\left(\mathcal{X}_{uvw}-\frac{\sqrt{2\mu}}{a_i}\right)^2 \mathds{1}_{\left\{\mathcal{X}_{uvw} \leq \frac{\sqrt{2 \mu  }}{a_i} \right\}}; \\
		G(\mathcal{X}) = & \  |\mathcal{X}|; \\
		X = & \ \{\mathcal{X} \ | \ \mathcal{X}_{uvw} \geq 0 \text{ for all } u,v,w  \}.
	\end{split}
\end{equation*}
Here $a_i = \|\mathcal{T}(\mathcal{A} \ast \delta_{uvw} )\|_F$ and $i = (w-1)mn+(v-1)m+u. $
The algorithm is summarized as Algorithm~\ref{alg:outer}. 
\begin{algorithm}[h]
\begin{algorithmic}[1]
\Require $\mathcal{X}^{(0)}\in \mathds{R}^{m\times n \times d}$ and $G\in \mathds{R}^{m \times n}$. Set $ \mu$.
\Ensure The solution $\mathcal{X}^\ast$ which is the minimizer in the last outer iteration.
\Repeat
    \State Compute $W^{(k)}_{uvw} = \left( a_i \sqrt{2\mu} - a_i^2\mathcal{X}^{(k)}_{uvw} \right) \mathds{1}_{\left\{\mathcal{X}^{(k)}_{uvw} \leq \frac{\sqrt{2 \mu  }}{a_i} \right\}};$
    \State Given $G$,  $W^{(k)}_{uvw}$,  obtain $\mathcal{X}^{(k)}$ by solving
    \begin{equation}
        \mathcal{X}^{(k+1)} = \argmin_{\mathcal{X}\geq 0  } \left\{ \frac{1}{2} \left\|\mathcal{T}(\mathcal{A} \ast \mathcal{X} )+b 1 -G\right\|_F^2 + \sum_{u,v,w=1}^{m,n,d} W^{(k)}_{uvw}  |\mathcal{X}_{uvw} | \right\}.\label{equ:subproblem}
    \end{equation}
   \Until{convergence}
\end{algorithmic}
\caption{Iterative reweighted $\ell_1$ algorithm (IRL1) for the rotating PSF problem}
\label{alg:outer}
\end{algorithm}

{\bf Remark: }The minimization problem in \eqref{equ:subproblem} of IRL1 is a weighted  $\ell_1$ model with nonnegative constraints. In  \cite{Rice2016generalized}, the $\ell_1$ model ``without'' nonnegative constraints is solved by the alternating direction method of multipliers (ADMM).

\section{Numerical Results}\label{sec:numerical}

In this section, we apply our optimization  approach to solving  simulated rotating PSF problems for point source localization and compare it to some other competing optimization methods. The codes of our algorithm and the others with which we compared our method were written in $\mathrm{MATLAB \ 9.0 \ (R2016a)}, $ and all our numerical experiments were conducted on a typical personal computer with a standard CPU (Intel i7-6700, 3.4GHz). 

The fidelity of localization is assessed in terms of the {\bf recall rate}, defined as the {\it ratio of the number of identified true positive point sources over the number of true positive point sources}, and the {\bf precision rate}, defined as the {\it ratio of the number of identified true positive point sources over the number of all point sources} obtained by the algorithm; see \cite{Book_SR_micro2017}. 

To distinguish  true positives from false positives for the estimated point sources, we need to determine the minimum total distance between them and the true point sources. Here all 2D simulated observed images are described by 96-by-96 matrices. We set the number of zones of the spiral phase mask responsible for the rotating PSF at $L=7$  and the aperture-plane side length as 4 which sets the pixel resolution in the 2D image (FFT) plane as 1/4 in units of $\lambda z_I/R$. 
The dictionary corresponding to our discretized 3D space contains 21 slices in the axial direction, with the corresponding values of the defocus parameter, $\zeta$, distributed uniformly over the range, $[-21, \  21]$.  
According to the Abbe-Rayleigh resolution criterion,  two point sources that are within $(1/2)\lambda z_I/R$ of each other and lying in the same transverse plane cannot be separated in the limit of low intensities.
In view of this criterion and our choice of the aperture-plane side length and if we assume conservatively that our algorithm does not yield any significant super-resolution, we must regard two point sources that  are within 2 image pixel units of each other as a single point source.
Analogously, two point sources along the same line of sight ({\it i.e.,} with the same $x,y$ coordinates) that are axially separated from each other within a single unit of $\zeta$  must also be regarded as a single point source. 
  
As in real problems, our simulation does not assume that the point sources are on the grid points. Rather, a number of point sources are randomly generated in a 3D continuous image space with certain fluxes. We consider a variety of source densities, from 5 point sources to 40 point sources in the same size space.  For each case, we randomly generate 20 observed images and use them for training the parameters in our algorithm, and then test 50 simulated images with the well-selected parameters. 
 The number of photons emitted by each point source follows a Poisson distribution with a mean of 2000 photons. 


 For adding the Gaussian noise, we use the MATLAB command 
  \begin{equation*}
  	\verb|G = I0  + b + sigma*randn(Np)|,
  \end{equation*}
where $\verb|b|$ is the uniform background noise which we set to a typical value 5. Here, $\verb|I0|$ is the 2D original image formed by  adding all the images of the point sources without noise, and  $\verb|Np| = 96$ is the size of the images. The noise level is denoted as $\verb|sigma|$ and we choose  it  to be   10\% of the highest pixel value in original image $\verb|I0|$. Here, \verb|randn| is the $\mathrm{MATLAB}$ command for  the Gaussian distribution with the mean as 0 and standard deviation as 1. 

\begin{figure}[htbp]
\centering
\subfloat[Observed image]{\includegraphics[width=0.3\textwidth]{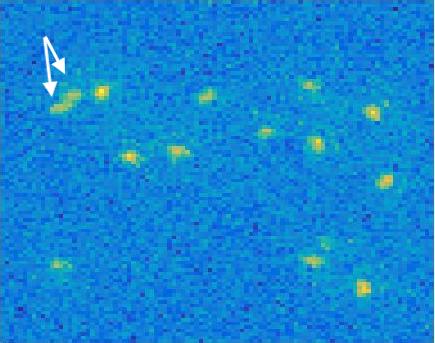}}
\hspace{0.01mm}
\subfloat[Estimated locations in 2D]{\includegraphics[width=0.3\textwidth]{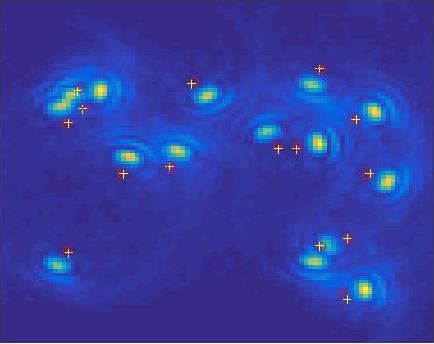}}
\hspace{0.01mm}
\subfloat[Estimated locations in 3D]{\includegraphics[width=0.3\textwidth]{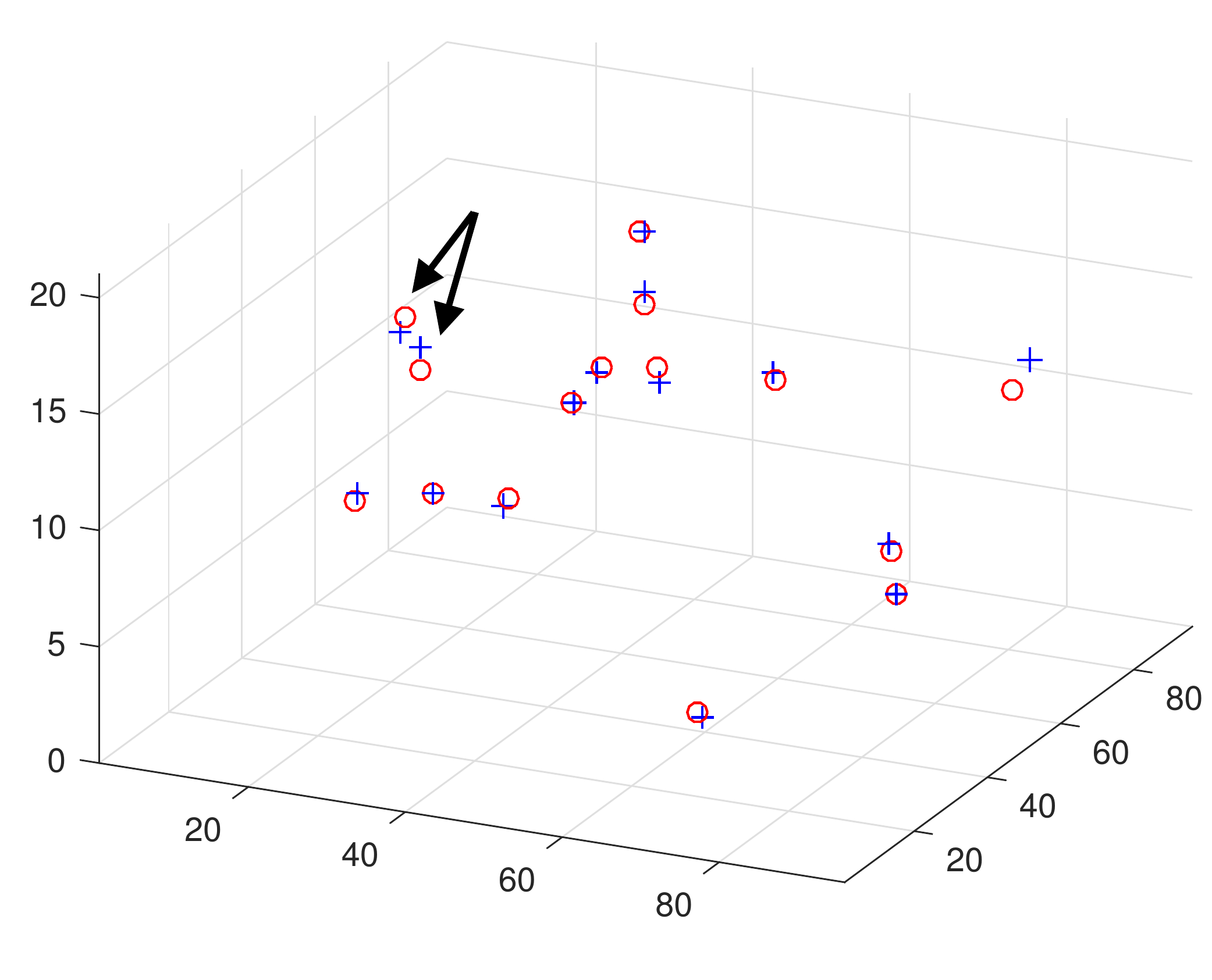}}
\caption{Localizations for the 15 point sources case. ``$\circ$'' denotes the location of the ground truth point source  and ``+''  the  location of the estimated point source. }\label{fig:15_cel0}
\end{figure}

\begin{figure}[htbp]
\centering
\subfloat[Observed image]{\includegraphics[width=0.3\textwidth]{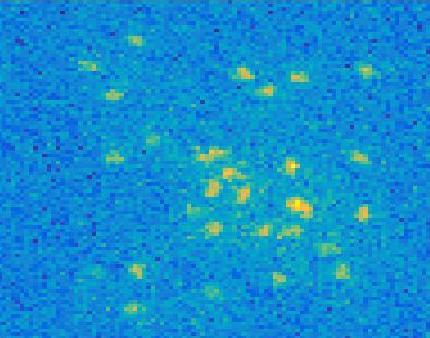}}
\hspace{0.01mm}
\subfloat[Estimated locations in 2D]{\includegraphics[width=0.3\textwidth]{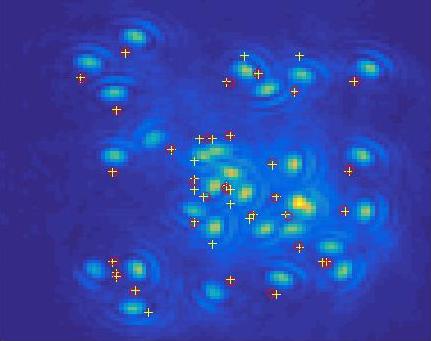}}
\hspace{0.01mm}
\subfloat[Estimated locations in 3D]{\includegraphics[width=0.3\textwidth]{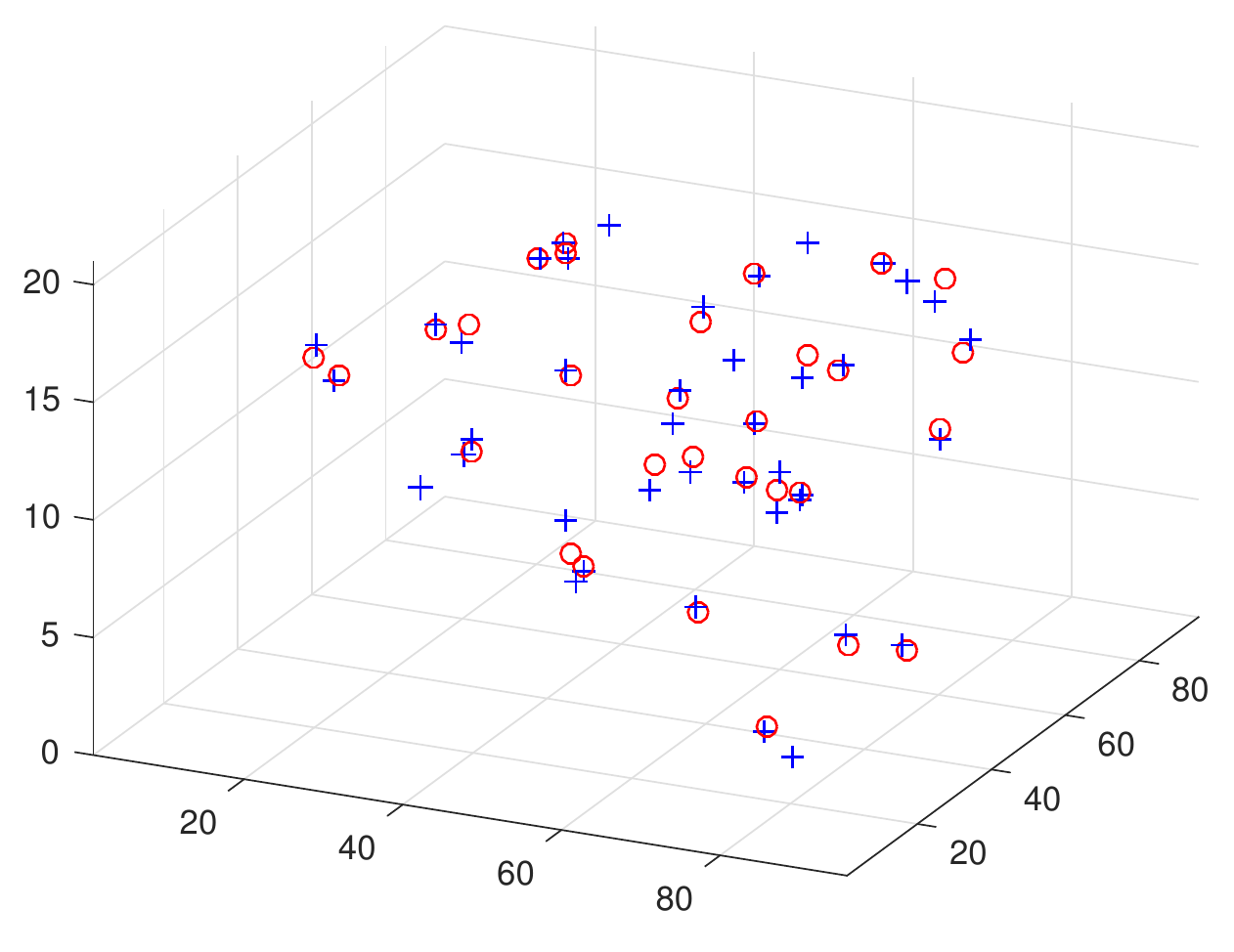}}
\caption{Localizations for the 30 point sources case. }\label{fig:30_cel0}
\end{figure}

We test our CEL0 based algorithm for several point-source densities. Figure~\ref{fig:15_cel0}-\ref{fig:30_cel0} consider examples of 15 point sources and 30 point sources, respectively.

From both observed images (see Figure~\ref{fig:15_cel0}(a) and Figure~\ref{fig:30_cel0}(a)), {neither the secondary rings nor the angle of rotation of the PSF are easily identifed}, which means we cannot use a calibration method \cite{huang2008cylendricalstorm,DH2009pavani}. In Figure~\ref{fig:15_cel0}(a), there are two overlapping PSF images corresponding to two different point sources, and our optimization approach is still able to distinguish and estimate them; see the arrows in Figure~\ref{fig:15_cel0}(a),(c).
  
In Figure~\ref{fig:30_cel0},  many PSF  images are overlapping corresponding to point sources that are very close. Our algorithm estimates the clusters of these point sources but {gives} more point sources than their ground-truth number. 
 

Next, we compare our algorithm with  two other regularization methods: an $\ell_1$ regularization model \cite{Rice2016generalized} as well as a new non-convex model \cite{mila2008nonconvex,mila2010nonconvex,tranformed2014minimization} called  transformed $\ell_1$ (TL1). We use $\ell_1$ to denote the $\ell_1$ regularization model whose regularization term is $$\mathcal{R}(\mathcal{X}):= \mu \left\| \mathcal{X} \right\|_1 = \mu \sum_{u,v,w = 1}^{m,n,d}|\mathcal{X}_{uvw}|. $$
Following \cite{Rice2016generalized}, we solve the optimization problem by ADMM. 
 For TL1, the regularization term is $$\mathcal{R}(\mathcal{X}):= \mu \sum_{u,v,w = 1}^{m,n,d} \theta(a;\mathcal{X}) = \mu \sum_{u,v,w = 1}^{m,n,d} \frac{|\mathcal{X}_{uvw}|}{a+|\mathcal{X}_{uvw}|}, $$
where $a$ is fixed parameter and determines the degree of non-convexity. We use IRL1 to solve this model with a similar scheme as Figure~\ref{alg:outer}. The only difference is $$W_{uvw}^{(k)} = \partial F_2(\mathcal{X}^{(k)}_{uvw}) = \frac{a\mu}{\left(a+\mathcal{X}^{(k)}_{uvw}\right)^2}, $$
where $F_2(\mathcal{Y}) = \mu \sum\limits_{u,v,w=1}^{m,n,d} \frac{\mathcal{Y}_{uvw}}{a+\mathcal{Y}_{uvw}}$. 

\begin{figure}[htbp]
\centering
\subfloat[$\ell_1$]{\includegraphics[width=0.32\textwidth, height=0.22\textheight]{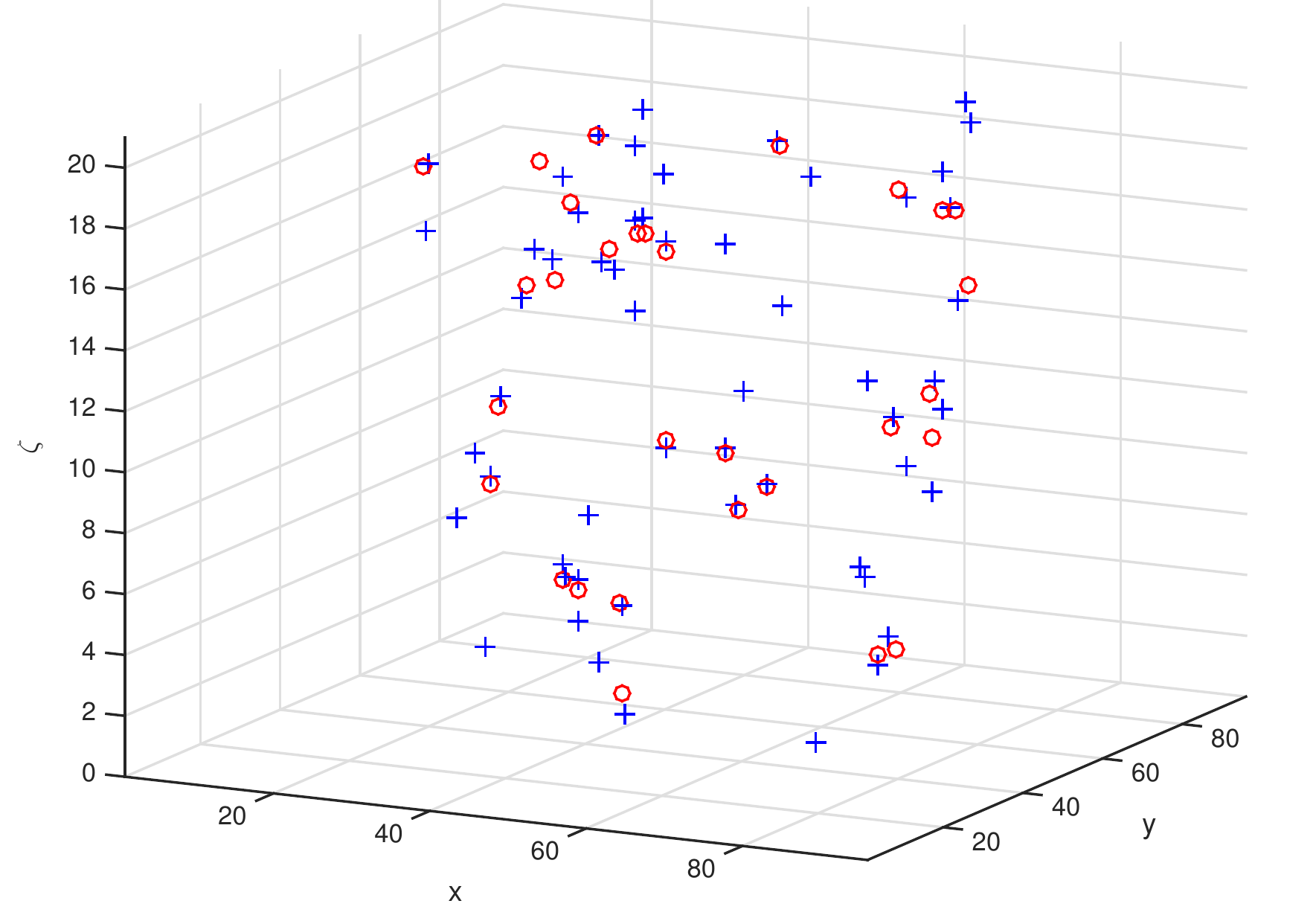}}
\subfloat[TL1]{\includegraphics[width=0.32\textwidth, height=0.22\textheight]{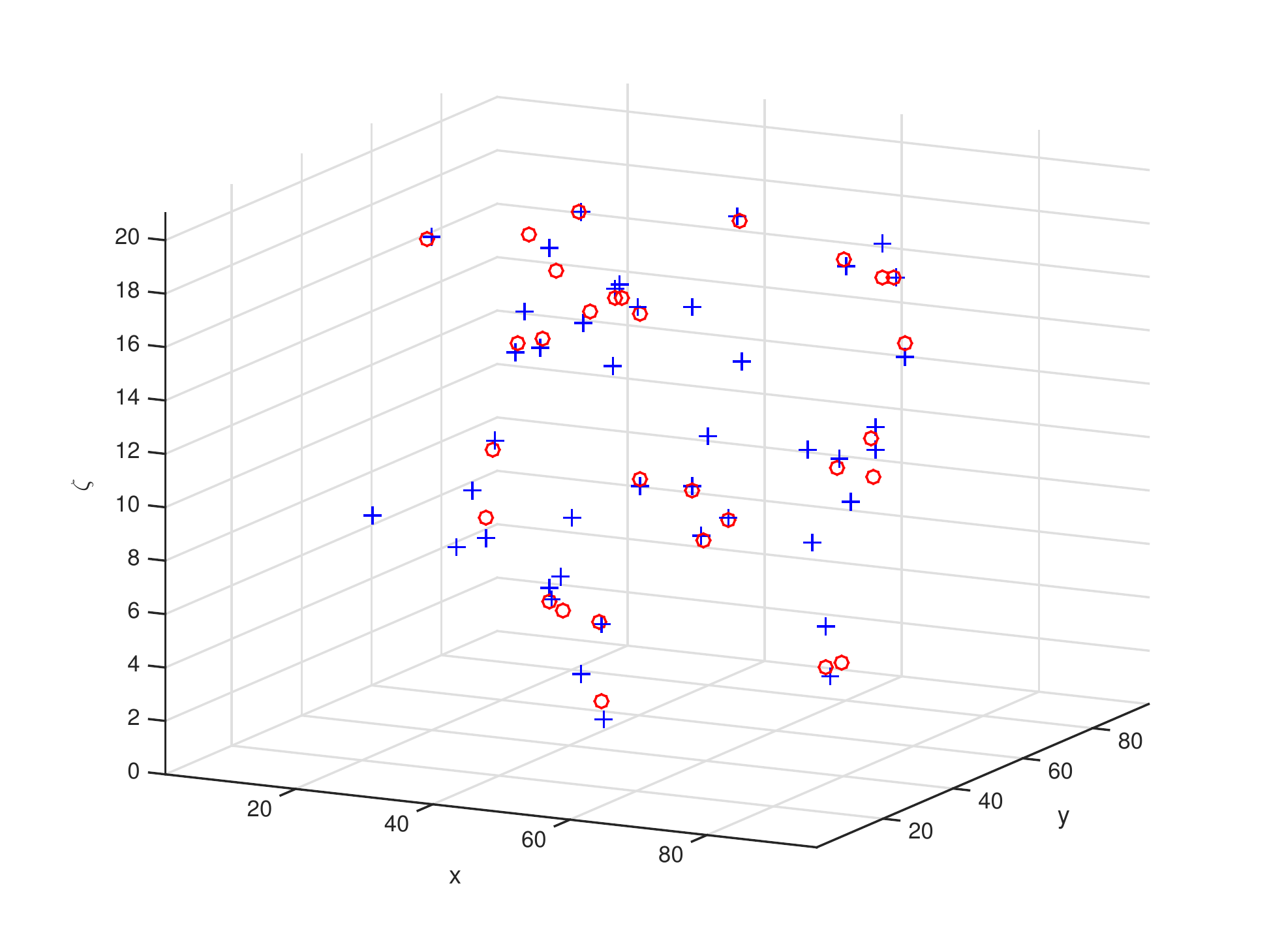}}
\subfloat[CEL0]{\includegraphics[width=0.32\textwidth, height=0.22\textheight]{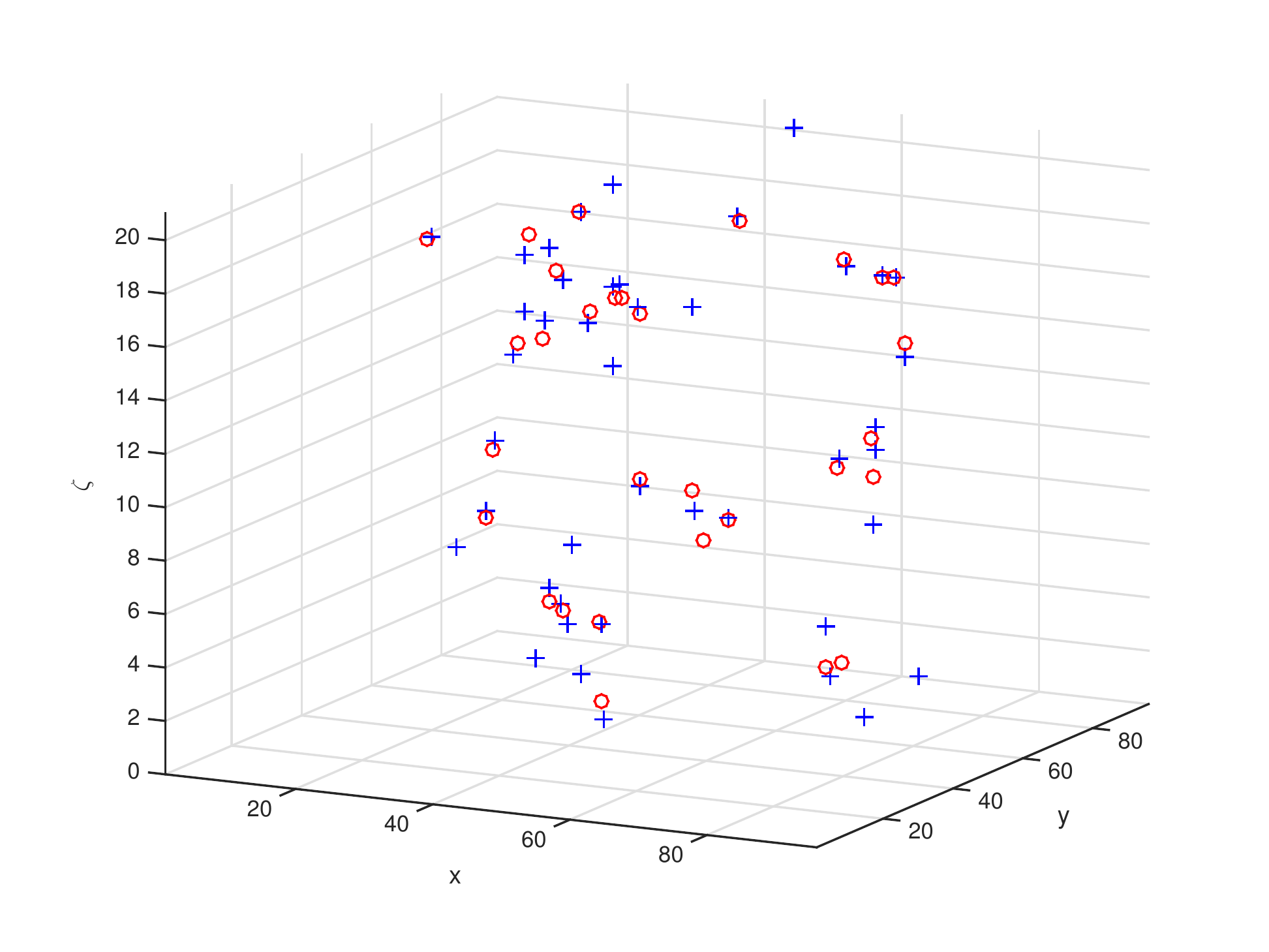}}
\caption{Localizations from three regularization models (30 point sources).  } 
\label{fig:compare_distr_30 cel0}
\end{figure}

In Figure~\ref{fig:compare_distr_30 cel0},  we again consider the 30 point sources case.  We see that $\ell_2$-$\ell_1$ has more false positives than other algorithms although it detects all the ground truth point sources. TL1 and our algorithm have different but fewer false positives.  

For further comparisons, we tested 50 different random images and computed the average of recall and precision rates in each density case for both algorithms; see 
Figure~\ref{Tab: cel0}. 

\begin{table}[htbp]
\begin{center}
\caption{Comparisons of  $\ell_2$-$\ell_1$ with our $\ell_2$-CEL0.  All the results are with post-processing.} 
\begin{tabular}{c|cc|cc|cc}
\hline  & \multicolumn{2}{c}{$\ell_1$} & \multicolumn{2}{c}{TL1} & \multicolumn{2}{c}{CEL0}\\
\hline No.  Sources & Recall &  Prec. & Recall &  Prec.   & Recall & Prec.   \\
\hline 10  & 94.80\% & 64.04\% & 89.60\% & 68.79\% & {\bf95.80}\% & {\bf79.72}\% \\
\hline 15  & 90.80\%& 61.68\%  & 87.07\% & 64.67\% &{\bf 93.20}\% &{\bf77.68}\% \\
\hline 20  & 86.60\%& 57.72\% & 83.30\% & 60.78\% & {\bf 89.30}\% &{\bf72.12}\% \\
\hline 30  & {\bf88.80}\% & 47.51\% & 79.80\% & 56.06\% &87.20\% &{\bf58.77}\%  \\
\hline 40  &  {\bf 81.50}\% & 42.03\% & 71.15\% & 48.81\% & 77.40\% &{\bf52.87}\%  \\
\hline
\end{tabular}\label{Tab: cel0}
\medskip
\end{center}
\end{table}

In Figure~\ref{Tab: cel0}, we see that our algorithm is better than $\ell_1$ and TL1  for almost all cases especially in precision rates. For example, in the cases of 10 and 15 point sources, the precision rate in our algorithm is over 10\% higher than the one in $\ell_1$.  In  the higher-density cases, like those with 30 and 40 sources, all methods have more than 5 false positives. We were able to mitigate the latter by further post-processing based on machine learning techniques, as in \cite{Rice2016generalized}. We set the maximum number of iterations for $\ell_1$ at 800, which guaranteed its convergence, and for CEL0 regularization and TL1 we set the maximum number of inner and outer iterations at 400 and 2, respectively.
 Here we  emphasize the advantage of our algorithm in providing a better initial guess than $\ell_1$ and TL1 with a similar cost time.

\section{Conclusions and Future Work}\label{sec:Conclusions}
We have proposed an optimization algorithm, based on a CEL0 penalty term, for the 3D localization of a swarm of randomly spaced point sources using a rotating PSF which has a single lobe in the image of each point source. This has distinct advantages over a double-lobe rotating PSF, e.g.  \cite{moerner2015single,DH2008pavani,DH2009pavani,Rice2016generalized}, especially in cases where the point source density is high and the photon number per source is small. This research focuses on the Gaussian noise case which describes  conventional CCD sensors in low per-pixel photon fluxes and large read-out noise. We note that at high source densities, the optimization can lead to false positives.

We employed a post-processing step based both on centroiding the locations of recovered sources that are tightly clustered  and thresholding the recovered flux values to eliminate obvious false positives from our recovery sources.  These techniques can be applied to other rotating PSFs as well as other depth-encoding PSFs for accurate 3D localization and flux recovery of point sources in a scene from its image data under the Poisson noise model.  Applications include not only 3D localization of  space debris, but also super-resolution 3D single-molecule localization microscopy, e.g.  \cite{Book_SR_micro2017,3dsml2017review}. 

Applying recently developed machine learning techniques for removing false positives instead of logistic regression models \cite{Rice2016generalized} will be considered. Tests of this algorithm based on real data collected using phase masks fabricated for both applications are currently being planned.  In addition,  work involving
snapshot multi-spectral/hyperspectral \cite{KK2020hypersp} imaging, which will permit accurate  material characterization, as well as higher 3D
resolution and localization of space microdebris via a sequence of snapshots is under way.

\bibliographystyle{plain}
\bibliography{ref_rPSF}

\end{document}